\documentclass[fleqn,usenatbib]{mnras}
\usepackage{newtxtext,newtxmath}
\usepackage[T1]{fontenc}
\usepackage{ae,aecompl}
\usepackage{amsmath}
\usepackage{graphicx}
\usepackage[mathscr]{euscript}

\let\mathscr\relax % so we can load this and rsfs
\usepackage[scr]{rsfso}
\usepackage{enumitem}
\usepackage{tabularx, booktabs, makecell}
\usepackage{siunitx}
\usepackage{rotating}
\usepackage{adjustbox}
\usepackage{array}
\newcommand{\be}{\begin{equation}}
\newcommand{\ee}{\end{equation}}
\newcommand{\daa}{\Delta\alpha/\alpha}

\DeclareMathAlphabet{\mathpzc}{OT1}{pzc}{m}{it}
\hypersetup{pdfauthor={Name}}

\title[Blinding and bias]{Varying alpha, blinding, and bias in existing measurements}

\author[Lee et al]{Chung-Chi Lee$^1$,
John K. Webb$^1$,
Robert F. Carswell$^2$,
Vladimir A. Dzuba$^3$,
Victor V. Flambaum$^3$, \newauthor
Dinko Milakovi{\'c}$^{\,4,5,6}$.
\\\\
$^1$Clare Hall, University of Cambridge, Herschel Rd, Cambridge CB3 9AL.\\
$^2$Institute of Astronomy, University of Cambridge, Madingley Road, Cambridge CB3 0HA, UK.\\
$^3$School of Physics, University of New South Wales, Sydney, NSW 2052, Australia.\\
$^4$Institute for Fundamental Physics of the Universe, Via Beirut, 2, I-34151, Grignano, Italy.\\
$^5$INAF -- Osservatorio Astronomico di Trieste, via Tiepolo 11, I-34131, Trieste, Italy.\\
$^6$INFN, Sezione di Trieste, Via Bonomea 265, I-34136, Trieste, Italy.
}

\date{Accepted XXX. Received YYY; in original form ZZZ}
\pubyear{2022}

\begin{document}
\label{firstpage}
\pagerange{\pageref{firstpage}--\pageref{lastpage}}
\maketitle

\begin{abstract}
The high resolution spectrograph ESPRESSO on the VLT allows measurements of fundamental constants at unprecedented precision and hence enables tests for spacetime variations predicted by some theories. In a series of recent papers, we developed optimal analysis procedures that both exposes and eliminates the subjectivity and bias in previous quasar absorption system measurements. In this paper we analyse the ESPRESSO spectrum of the absorption system at $z_{abs}=1.15$ towards the quasar HE0515-4414. Our goal here is not to provide a new unbiased measurement of $\daa$ in this system (that will be done separately). Rather, it is to carefully examine the impact of blinding procedures applied in the recent analysis of the same data by \cite{Murphy2022} (M22) and prior to that, in several other analyses. To do this we use supercomputer Monte Carlo AI calculations to generate a large number of independently constructed models of the absorption complex. Each model is obtained using {\sc ai-vpfit}, with $\daa$ fixed until a ``final'' model is obtained, at which point $\daa$ is then released as a free parameter for one final optimisation. The results show that the ``measured'' value of $\daa$ is systematically biased towards the initially-fixed value i.e. this process produces meaningless measurements. The implication is straightforward: to avoid bias, all future measurements must include $\daa$ as a free parameter from the beginning of the modelling process.
\end{abstract}


\section{Introduction} \label{sec:Intro}

Blinding methods are widely used, to great effect, across many scientific disciplines. \cite{Harrison2002, Roodman2003, Klein2005, Maccoun2015} give example applications in particle physics. \cite{Muir2020} give a comprehensive description of blinding in cosmology. These papers and many others show that blinding methods can guard against human bias. Nevertheless, in applying blinding techniques, one must be certain that doing so has no impact on the final measurement. 

The subject of the present paper, a particularly challenging measurement in cosmology, is searching for spacetime variations of the fine structure constant at high redshift. Theoretical motivations are diverse and many, a selection being those in \cite{Barrow2003, Barrow2012, Sola2015, Stadnik2015a, Stadnik2015b, Davoudiasl2019, Barros2022} with reviews provided by \cite{Uzan2011, Martins2017}. In this context, we recently used spectral simulations \citep{Webb2022} to emulate a blinding method most recently used in \cite{Murphy2022}, hereafter referred to as M22. This kind of method has been applied in many previous papers, the intended goal being to try and eliminate any possible human subjectivity in the analysis that could potentially emulate a non-zero measurement of $\daa = (\alpha_z - \alpha_0)/\alpha_0$, where the subscripts $z, 0$ indicate redshift and the terrestrial value, and $\alpha_{SI} = e^2/4\pi\epsilon_0\hbar c$. Nevertheless, a simple consideration suggests that the method could instead bias measurements towards zero; fixing $\daa=0$ amounts to forcing all rest-frame wavelengths in the initial modelling to be terrestrial values \citep{Webb2022}. If the true $\daa$ is not zero, forcing $\daa=0$ whilst the fit is developed necessarily results in a flawed model. This would not be a problem provided that the $\chi^2$-$\daa$ space is smooth, with a single minimum, since subsequently releasing $\daa$ as a free parameter, after the $\daa=0$ model has been obtained, should result in further iterations that reach the correct solution. However, we now know that $\chi^2$-$\daa$ space is {\it not} necessarily smooth \citep{Lee2021}, particularly if the absorption system is complex and requires a large number of free parameters. Therefore, the final step of ``switching on'' $\daa$ as a free parameter may not allow the fit to get out of its false local minimum, potentially generating a systematically biased $\daa$ measurement.

Applying the same analyses that are used on real data to simple simulated spectra confirmed the effect just described; the blinding method used in M22, which we referred to as ``distortion blinding'' in \cite{Webb2022} (since it involves imposing a slight distortion of the spectral wavelength scale as well as imposing a fixed $\daa=0$ during the model building phase of the analysis) was found to create a strong bias towards a null result. However, that analysis was indeed only carried out on simulated absorption systems, requiring confirmation using real data. In the present paper, we take that investigation further by extending the test to real data and more a complex absorption system, the same spectrum used in M22.

\section{The astronomical data} \label{sec:Data}

The analysis here is of the well known absorption complex at $z_{abs}=1.15$ towards the bright quasar HE0515-4414. The astronomical data were obtained using the high-resolution VLT spectrograph, ESPRESSO \citep[Echelle SPectrograph for Rocky Exoplanet and Stable Spectroscopic Observations, ][]{espresso2021} . The spectral resolving power is $R=\lambda/\Delta\lambda\sim145,000$, the signal to noise per 0.4 km/s pixel is approximately 105 (at 6000\AA) and 85 (at 5000\AA). 

These data have been described in detail in M22 and the extracted and calibrated spectra are publicly available\footnote{The ESPRESSO spectrum of HE0515-4414 used in this paper and in M22 is available at \url{https://doi.org/10.5281/zenodo.5512490}}. This absorption system was also studied in detail recently by \cite{Milakovic2021} (using different observational data). Those data were also high resolution ($R=115,000$), with an average signal to noise $\sim$50 per 0.83 km/s pixel. 

Most importantly, wavelength calibration of the ESPRESSO data was done using a laser frequency comb (LFC), so wavelength calibration uncertainties are negligible and can be ignored in the context of varying alpha measurements. LFC line profiles have been explored in \cite{DinkoPhD, Zhao2021} and used to reveal a non-Gaussian instrumental profile (IP) for the European Southern Observatory's HARPS instrument. That study prompted similar investigation using ESPRESSO LFC lines, which are also non-Gaussian \citep{Schmidt2021}. For a robust measurement of the fine structure constant, it is important to use the correct IP, or at least carefully examine the impact of assuming a Gaussian profile. However since in this paper we are concerned only with assessing the impact of blinding methods that have been applied in some previous analyses, we adopt a Gaussian for convolving theoretical absorption line models with the instrumental profile, enabling a direct comparison with M22's model which assumed a Gaussian IP. 

In this study, we do not use the entire $z=1.15$ complex but instead use only region 1 \citep{Milakovic2021}. The work described here is computationally demanding, region 2 does not constrain $\daa$ well, and region 3 is more complex than region 1 and would take considerably longer to compute. Moreover, the results from region 1 alone yield clear conclusions.

\section{Revealing the bias caused by blinding} \label{sec:blind}

The M22 blinding process comprises two stages. Firstly, both long-range and intra-order distortions of the wavelength scale are applied to each exposure. It is claimed that the net effect of these distortions are such that there is an (artificial) $\daa$ added to the data of up to $|\daa| \le 5 \times 10^{-6}$. Secondly, M22 use {\sc vpfit} to model the distorted spectrum, with $\daa$ fixed at zero throughout the model building process, and allowed to vary only once the final velocity structure has been derived. In the tests carried out here, we do not attempt to emulate the first part of this process because the quantitative details of these distortions are not given in M22 and have not been published elsewhere as far as we know. Therefore we investigate only the second aspect -- the impact of initially-fixed $\daa$ on the final measurement.

\subsection{Methods}

The following procedures were used:
\begin{enumerate}[wide, labelwidth=!,itemindent=!,labelindent=0pt, leftmargin=0em, parsep=0pt]
\item {\sc ai-vpfit} \citep{Lee2020AI-VPFIT} and the Many Multiplet Method were used to obtain best-fit models, initially using fixed $\daa$. Redshifted wavelengths, for each fixed $\daa$, were calculated using laboratory rest-frame wavelengths (Section \ref{sec:atomicparams}) and sensitivity coefficients describing wavelength shifts (q-coefficients). Redshift wavelengths were derived using $\omega_z = \omega_0 + q \left(\alpha_z^2/\alpha_0^2 - 1 \right)$ where $\omega$ denotes transition frequency and the subscripts $z$ and $0$ indicate redshifted and terrestrial values. The Many Multiplet Method and calculations of sensitivity coefficients in an astronomical context were introduced in \cite{Dzuba1999a, Dzuba1999b, Webb1999}.
\item The {\sc ai-vpfit} {\it primary species} \citep[section 2, ][]{Lee2020AI-VPFIT} was MgII 2796. Modelling initially used 3 fixed values, $\daa = +10^{-5}$, $0$, and $-10^{-5}$. However, during the course of this work, we noticed a systematic tendency for the $\daa = 0$ and $-10^{-5}$ {\sc ai-vpfit} models to drift towards more positive $\daa$ (Section \ref{sec:Results}), prompting us to try a fourth fixed value, $\daa = +3 \times 10^{-5}$, to see if the effect persisted or changed.
\item Two line broadening mechanisms (turbulent and compound) (Section \ref{sec:Broadening}) and two information criteria (IC), AICc and SpiC, \cite{Webb2021}, were used.
\item After each final fixed $\daa$ model was obtained from {\sc ai-vpfit}, {\sc vpfit} was used, releasing $\daa$ as a free parameter (along with all other model parameters). We used two versions of {\sc vpfit}, v12.1, and as an additional check, a modified version of v12.2. The former uses numerical finite difference derivatives in calculating the Hessian matrix and gradient vector \citep{WebbVPFIT2021} whilst the latter uses analytic derivatives \citep{Lee2021Addendum}. Both gave consistent final results. This check removes any possibility that {\sc vpfit} could stop iterating too early due to a numerical accuracy effect. 
\end{enumerate}

Using the procedures above, a total of 400 absorption system models were generated (25 models for each of the 16 settings). Calculations were carried out using the OzSTAR supercomputer facility\footnote{https://www.swinburne.edu.au/research/facilities-equipment/supercomputer/}, requiring a total of $\sim$620,000 processor hours.

The {\sc ai-vpfit} and associated methodologies have been described in detail in other papers so we confine the discussion here to a brief summary. Artificial Intelligence procedures were developed, initially by \cite{gvpfit2017, Bainbridge2017}, subsequently by \cite{Lee2020AI-VPFIT}, to fully automate the modelling of high resolution spectra of quasar absorption systems. These methods require no human decisions during model construction. Quasar absorption systems have complex velocity structure with many components. The AI process randomly places trial components within the absorption complex, such that repeated {\sc ai-vpfit} calculations construct the model differently each time. This property allows us to map out the best-fit $\chi^2$ {\it vs.} parameter space, identifying any possible local minima i.e. model non-uniqueness \citep{Lee2021}. Each realisation of an {\sc ai-vpfit} model, because of the the different random placement each time, also emulates different interactive modellers. An information criterion is used to determine the number of free parameters used to model a complex \citep{Webb2021}. Voigt profile models are computed with high precision, to the machine precision of the computer used \citep{WebbVPFIT2021, Lee2021Addendum}. In applying these procedures, {\sc ai-vpfit} model generation is both ``blinded'' (because there is no human interaction) and unbiased (because no human interaction means no human bias and because candidate absorption components are placed randomly across a complex). Known systematic effects have been identified and quantified and are summarised in \cite{Webb2022}.

\subsection{Line broadening} \label{sec:Broadening}

The general (and physically appropriate) absorption line broadening model is compound broadening, such that the observed line width of an individual absorption component with atomic mass $m$ is given by
\be
b_{obs}^2 = b_{\textrm{turb}}^2 + \frac{2kT}{m}
\label{eq:btot}
\ee
However, in order to avoid including the cloud temperature $T$ as an additional free parameter, M22 (and many prior studies, including those involving two of the authors of the present paper) use turbulent broadening $(T=0)$. Whilst doing so may suggest that turbulent models require fewer free parameters than compound broadening models, in fact we now know that the opposite is true; additional absorption components are required to compensate for the poorer fit caused by the incorrect assumption of a single $b$ parameter for all species \cite{Webb2022}. Despite these line broadening considerations, our purpose here is not to create the most appropriate model for the $z_{abs}=1.15$ system towards the quasar HE0515-4414. Rather, the goal is to check on the ``distortion blinding'' approach employed in previous studies, most recently in M22. Therefore we apply {\it both} turbulent and compound broadening, presenting the results separately.

\subsection{Input atomic parameters and elemental isotopes} \label{sec:atomicparams}

{\sc vpfit} and {\sc ai-vpfit} require input atomic data (laboratory wavelengths, oscillator strengths, damping constants, etc., \cite{ascl:VPFIT2014, Lee2020AI-VPFIT}). The MgII isotopic wavelength spacings are of particular importance because they are reasonably well separated. If high redshift abundances differ from terrestrial values (as is expected, \cite{Kobayashi2020}), yet the observed profiles are modelled using terrestrial values, the inferred $\daa$ could be significantly biased \citep{Webb1999}. However, in this paper we are concerned only with assessing the impact of blinding methods that have been applied in some analyses, and in particular we want a direct comparison with the results in M22, so we use the default set of atomic parameters and isotope settings (terrestrial) supplied with {\sc vpfit}.

\section{Comparing ai-vpfit and M22 procedures} \label{sec:Comparison}

\subsection{The major difference: objectivity}
As described previously, {\sc ai-vpfit} model construction is carried out in this study by sequentially introducing randomly placed candidate absorption components and then allowing the non-linear least squares minimisation part of the code to iterate to a best fit. Model construction proceeds iteratively in this way. The final fit for each {\sc ai-vpfit} model is defined by an information criterion. Our analysis makes use of two ICs, the corrected Akaike Information Criterion (AICc) and the Spectral Information Criterion (SpIC) \cite{Webb2021}. Both ICs work well (SpIC is more suited to this application, but that is unimportant here) and there is no particular value of using two, other than comparison. Using an IC in general allows an optimal number of model parameters to be identified in an objective and reproducible way and is preferable to relying only on $\chi^2$. Discussions about the application of ICs in astrophysics are given in \cite{Liddle2004, Liddle2007} and a detailed technical treatment may be found in the book by \cite{Burnham2002}. Problems associated with noise characteristics in calculating ICs are described in \cite{Rossi2020}. Since $\chi^2$ asymptotes as more free parameters are introduced, it is relatively insensitive to the number of parameters chosen. Using $\chi^2$ normalised by the number of degrees of freedom to select an ``acceptable'' model is not sufficiently discriminatory because it requires the user to select a threshold normalised $\chi^2$ (which in practice is often vaguely defined to be ``around unity'') and because the spectral error array is notoriously hard to calculate accurately, so $\chi^2$ itself is only an approximation\footnote{The difficulties in accurately estimating the spectral error array are discussed in the {\sc rdgen} user guide \cite{web:VPFIT}.}. The general form of an IC is 
\be
\textrm{IC} = \chi^2 + \mathcal{P}(n_p,n_d)
\label{eq:IC}
\ee
where $\chi^2_{\nu} = \frac{1}{\nu} \sum_{i=1}^{n_d} \left[\left(d_i - f_i\right)/\sigma_i \right]^2$, $d_i$ is the spectral data array, $f_i$ is the model, $\sigma_i$ is the spectral error array, $n_d$ is the number of data points, $n_p$ is the total number of free parameters in the model, $\nu=n_d - n_p$ is the number of degrees of freedom, and $\chi^2 = \nu\chi^2_{\nu}$. The penalty term, $\mathcal{P}(n_p,n_d)$, increases with $n_p$, such that the IC minimises rather than asymptotes. ICs are also impacted by inaccurate spectral error arrays but nevertheless eliminate the user requirement to decide on what is or is not an acceptable $\chi^2$.

Modelling automation (i.e. the use of {\sc ai-vpfit} plus IC), removes all subjectivity and also permits Monte Carlo calculations, repeating the modelling process multiple times (different random seeds are used at each run) in order to map out $\daa$-IC space, hence revealing multiple minima should they exist. In contrast, the M22 model was selected according to the overall value of the normalised value of $\chi^2$ for the fit and by visually inspecting normalised residuals between model and data. This, and the human interactive nature of the model building process, means that the M22 model is subjective; a different human modelling the same data is likely to obtain a different model.

\subsection{The second difference: choice of free and fixed parameters}
The approach taken in using {\sc ai-vpfit} is that all species are assumed to exist in every redshift component within the absorption complex. This is physically correct and should not be thought of as an ``assumption'' or an approximation since every column density of every redshift component is a free fitting parameter which can iterate to a negligible value if the data require it. In some redshift components, column densities of weak components/species may fall below detection thresholds, even at the high signal to noise and high resolution of the ESPRESSO data used in this analysis. To deal with this numerically, we set a minimum column density threshold (for any species) of $\log N = 7.99$. Whilst superficially this method appears to contrast with that of M22, in fact the two approaches are similar (in this respect only) because a column density of $\log N = 7.99$ ($N$ being measured in atoms\,cm$^{-2}$) is far below any realistic detection threshold.

\subsection{The third difference: continuum and zero level parameters} \label{sec:Continua}

A further difference between our {\sc ai-vpfit} models and the M22 model is that the latter contains {\it fixed} continuum parameters. In the M22 model, the continuum normalisation is treated as a {\it fixed} parameter, (not at unity), with a fixed slope of zero. In the {\sc ai-vpfit} models, each spectral segment contains its own two linear continuum correction parameters\footnote{Normalisation and slope, see {\sc vpfit} user guide, \url{{https://people.ast.cam.ac.uk/~rfc/}}}. Fixing continuum parameters is generally inadvisable (and unjustified) because it is likely to artificially reduce the uncertainty on $\daa$ (although we have not attempted to quantify the effect and it may be negligible in this case). In the {\sc ai-vpfit} models, we include 2 additional free continuum parameters (normalisation and slope) in each spectral segment used in the fitting process. 

The zero level in each spectral segment also has some uncertainty and it can be important to allow for this when modelling. However, this parameter need not necessarily be well constrained unless the spectral region contains at least one well-saturated absorption line. In this case, the data does not. Therefore we simply emulate the M22 model in this respect by including just one free parameter in the MgII 2796 spectral region, the strongest line closest to saturation.

\subsection{Number of parameters in the M22 model}

We summarise the properties of the M22 model here, for comparison with the {\sc ai-vpfit} models described in Section \ref{sec:Results}. There are 41 MgII components in the M22 model, of which 26 exhibit detectable FeII absorption and 27 exhibit detectable MgI absorption. By ``detectable'', we do not mean above some statistical significance level, but rather that a component has been included in the interactively derived absorption system model. Of the 41 MgII components, 21 velocity components are detected in all 3 ions, 5 components are detected only in MgII and FeII, 6 components are present only in MgII and MgI, so 9 components exhibit only MgII. The best-fit M22 model has a normalised chi squared of 0.795 and a total of 177 free parameters. These details are shown in Table \ref{tab:n_components} for comparison with the {\sc ai-vpfit} results given in Section \ref{sec:Results}.

\section{Results} \label{sec:Results}

\begin{figure*}
\centering
\includegraphics[width=0.75\linewidth]{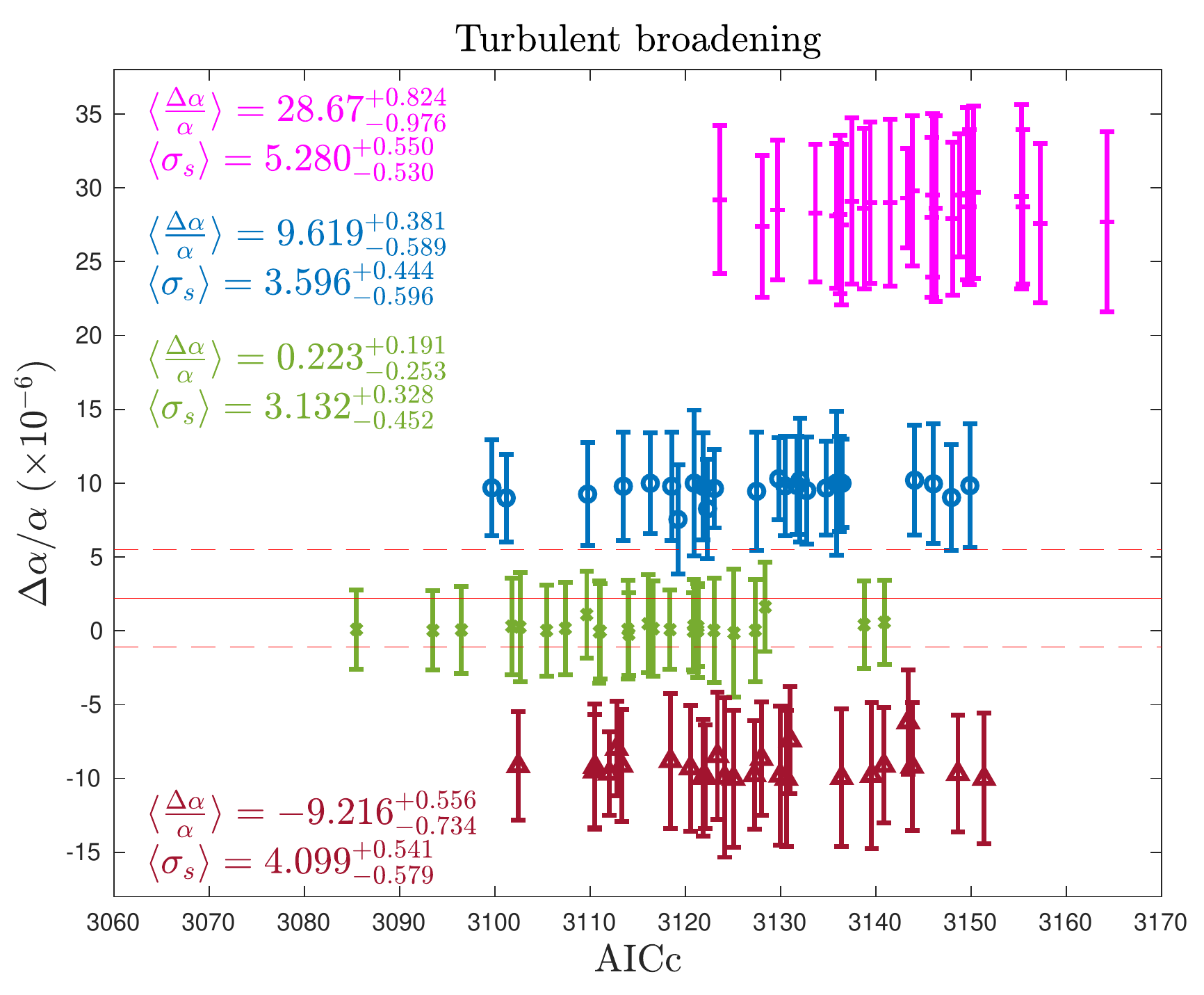}
\includegraphics[width=0.75\linewidth]{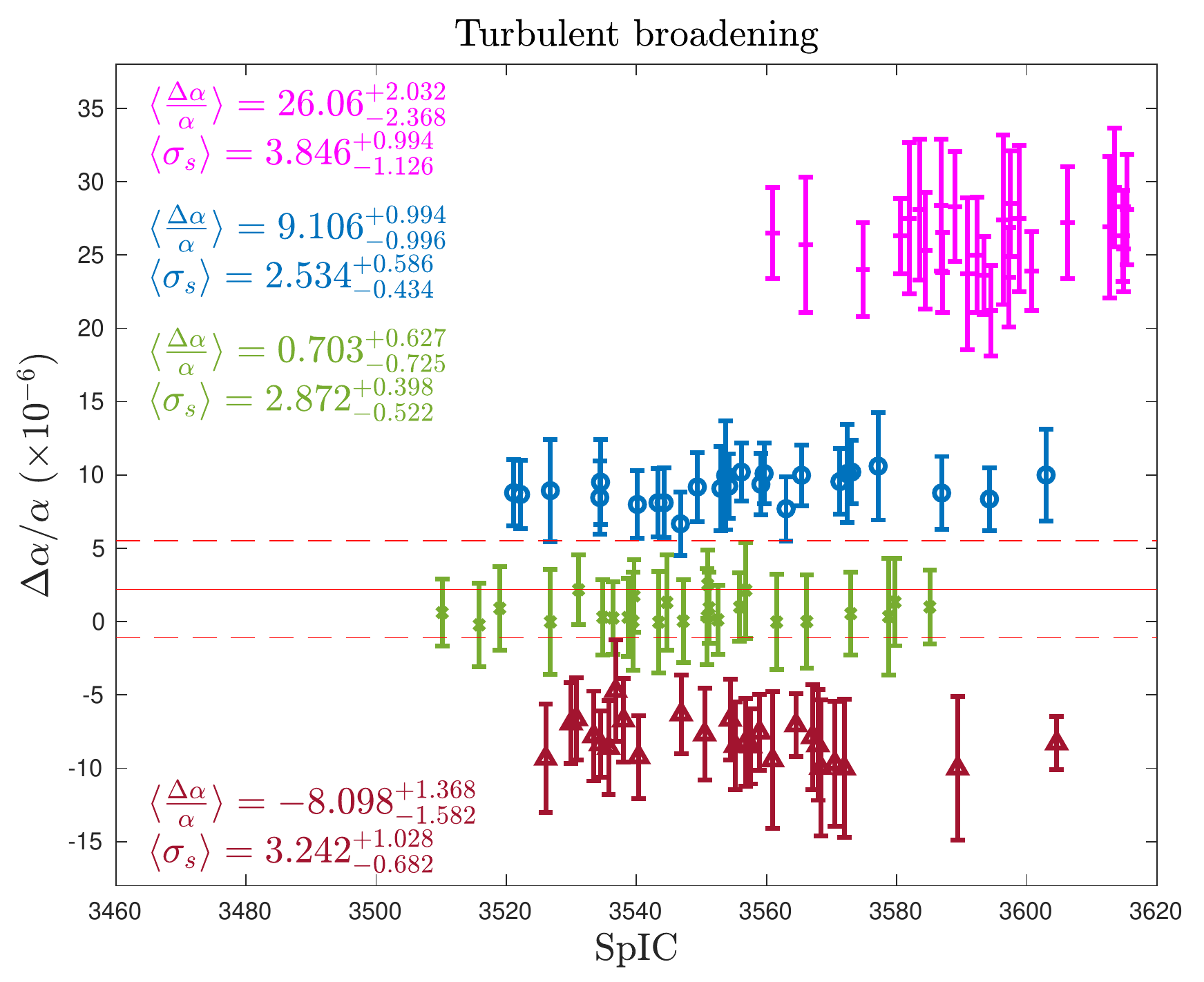}
\caption{Four sets of 25 {\sc ai-vpfit} models using turbulent line broadening (Equation \ref{eq:btot} with $T=0$). Top panel: AICc (numerical values are given in Table \ref{tab:aicctb_results}). Lower panel: SpIC (numerical values are given in Table \ref{tab:spictb_results}). The four sets (from top to bottom) correspond to fixed input $\daa = +30$, $+10$, $0$, and $-10 \times 10^{-6}$. The horizontal lines illustrate the final M22 $\daa$ value and $\pm$1$\sigma$ uncertainty. Each of the 100 points illustrates the {\sc vpfit} error bar. Four numerical insets show the means of 25 $\daa$ values with the empirical scatter in each case, and the means of the corresponding 25 {\sc vpfit} uncertainties, also with the empirical scatter.
\label{fig:turbulent_ICs} 
}
\end{figure*}

\begin{figure*}
\centering
\includegraphics[width=0.75\linewidth]{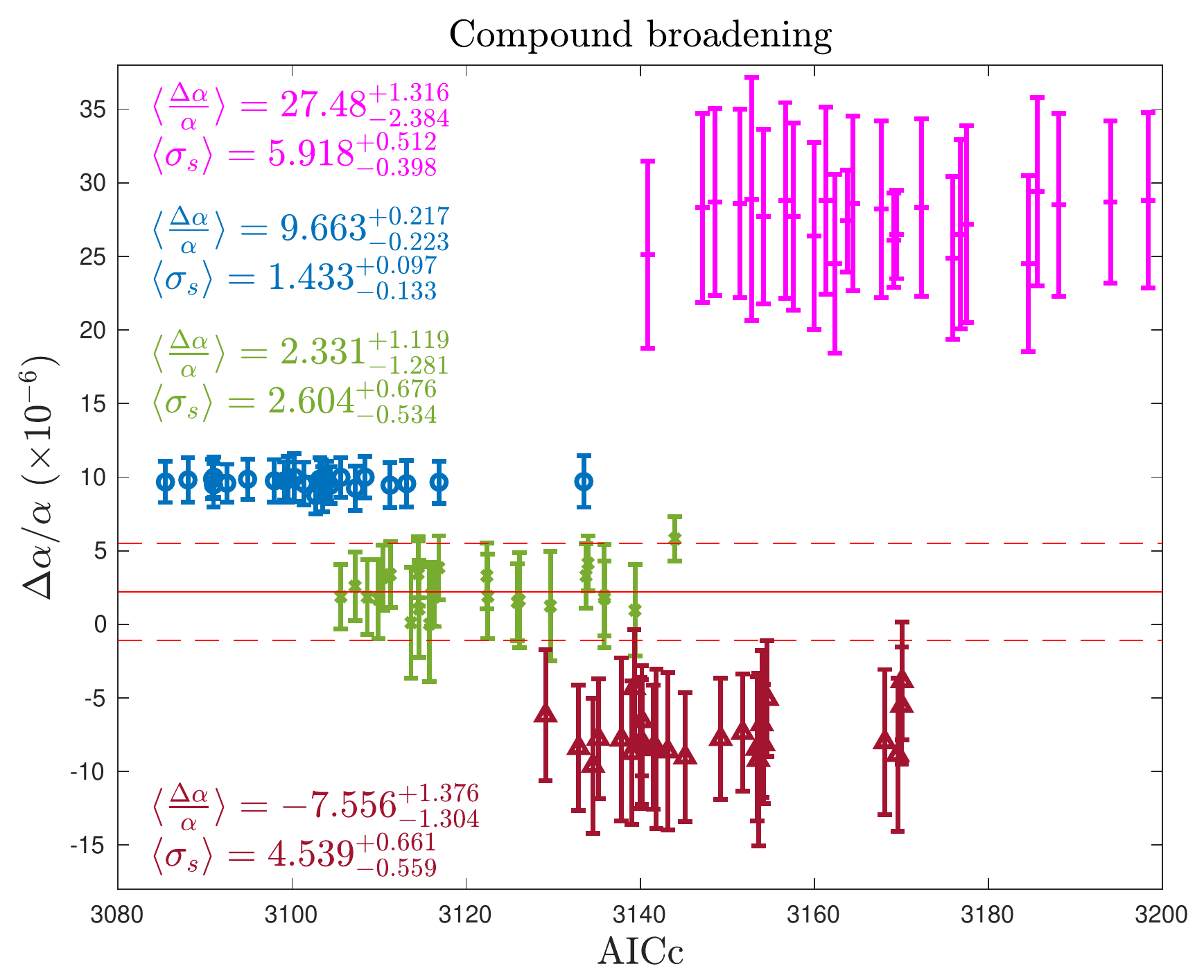}
\includegraphics[width=0.75\linewidth]{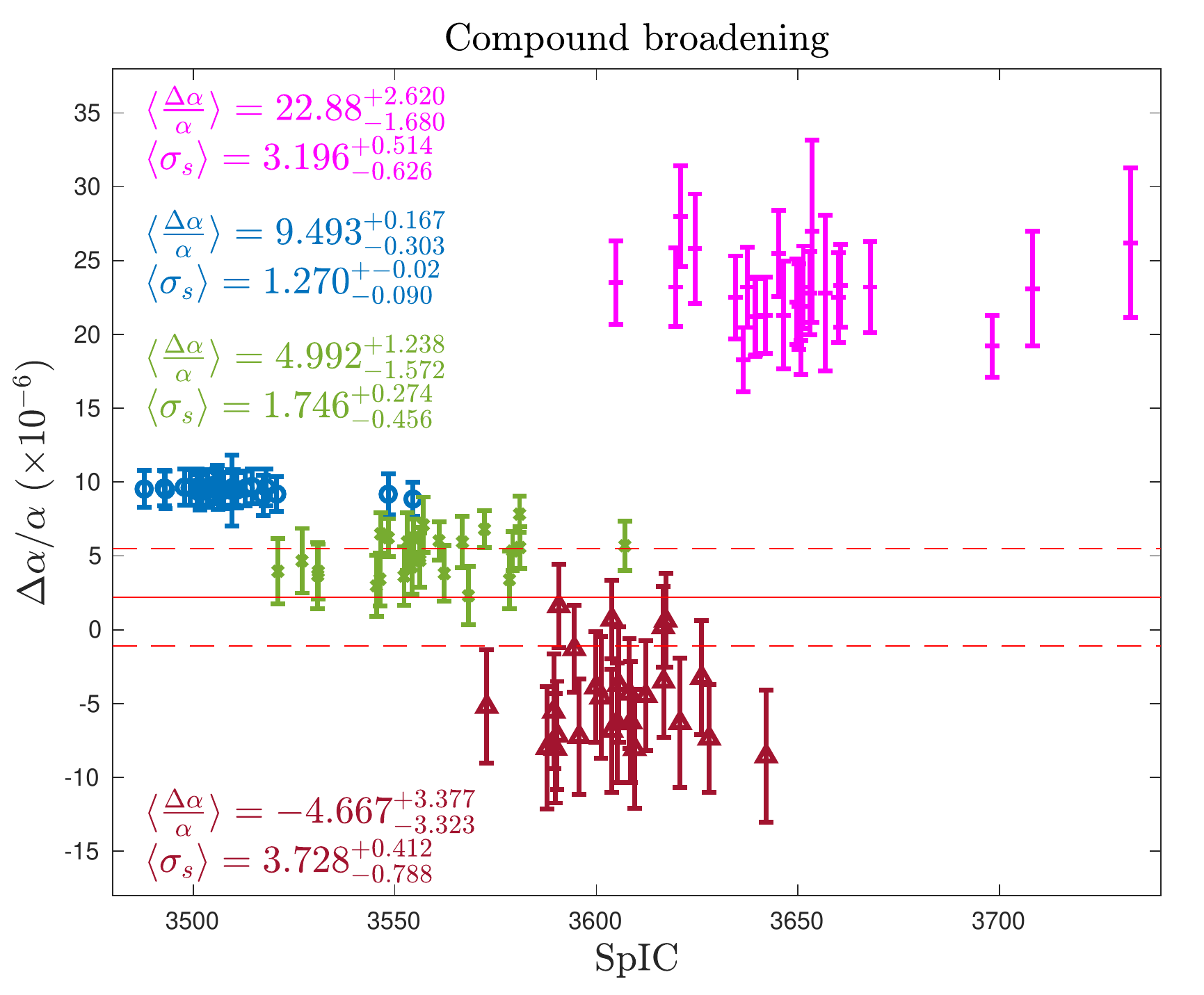}
\caption{As Figure \ref{fig:turbulent_ICs} but for compound broadening. Numerical values are given in Tables \ref{tab:aicccp_results} and \ref{tab:spiccp_results}.
\label{fig:compound_ICs} 
}
\end{figure*}

\begin{figure*}
\centering
\includegraphics[width=0.9\linewidth]{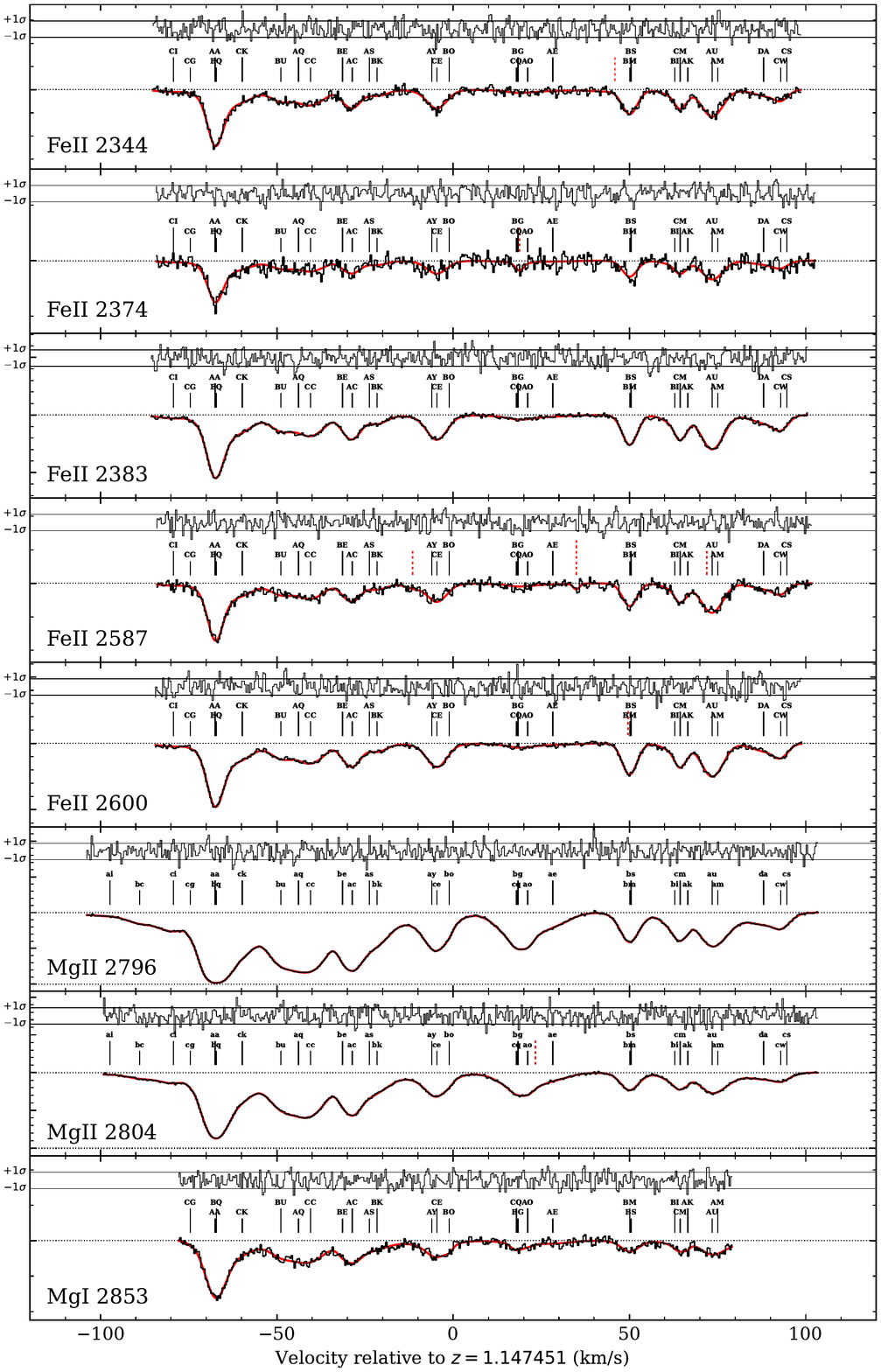}
\caption{Example {\sc ai-vpfit} model, with fit parameters: turbulent broadening, AICc, $\daa=0$ fixed until final iteration (see Section \ref{sec:Results} for details). The figure shows model 17 in the third column of Table \ref{tab:aicctb_results} (chosen because it has the lowest $N_{p}$ in that column). Black solid ticks are metals, red dashed ticks are interlopers. 
\label{fig:all} 
}
\end{figure*}

\begin{figure*}
\centering
\includegraphics[width=0.8\linewidth]{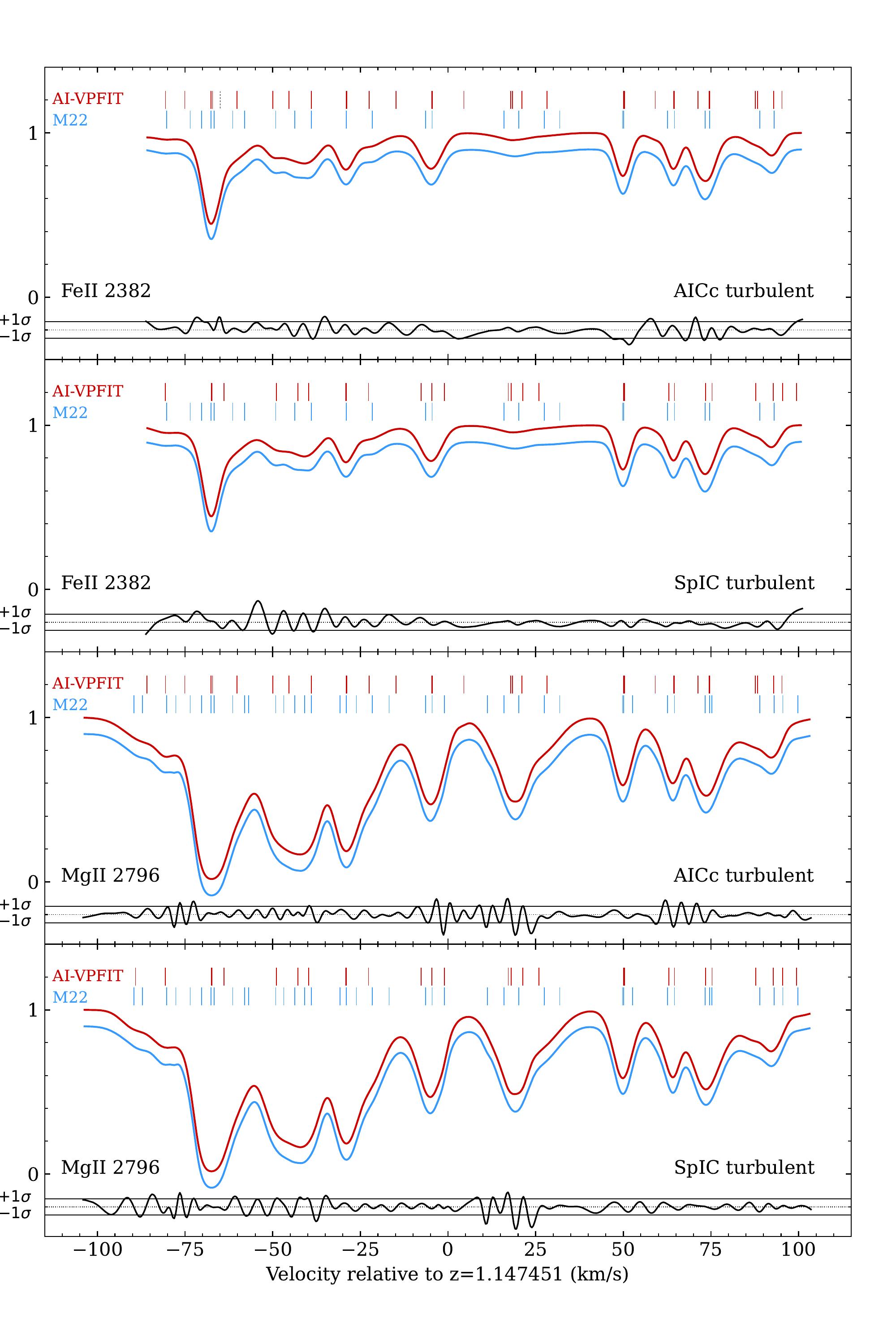}
\caption{Comparison between one {\sc ai-vpfit} model (red line) and the M22 model (blue line). The quasar spectrum is not shown. The blue curve is artificially offset from the red one by -10\% for illustrative purposes. The residuals underneath each panel show the difference between the models divided by the spectral error array. The horizontal lines either side of the normalised residuals illustrate the approximate $\pm$1$\sigma$ bounds. The two rows of tick marks show positions of individual components in the models. Solid coloured ticks mark metals and dotted black ticks mark interlopers. The legends within each panel provide further information. The models plotted are initially fixed $\daa=0$, AICc, turbulent, model 3 in Table \ref{tab:aicctb_results} and SpIC, turbulent, model number 3 in Table \ref{tab:spictb_results} (model number 3 was chosen at random).
\label{fig:residuals}
}
\end{figure*}

\begin{table*}
\centering
\begin{tabular}{| c m{2em} | m{2em} m{2em} m{2em} m{2em} m{2em} m{2em} m{4em} |}
\hline
Broadening & IC & MgII & MgI & FeII & $\langle\chi^2_{\nu}\rangle$ & $\langle$Int$\rangle$ & $\langle N_p \rangle$ & $\langle \daa \rangle$  \\
\hline \hline
\multicolumn{9}{| c |} {\bf{Input $\daa=+30 \times 10^{-6}$}} \\
\hline
Turbulent & AICc SpIC & 32.8 28.8 & 24.2 21.2 & 29.3 25.8 & 0.761 0.793 & 11.1  3.0 & 211.3 168.3 & +28.7 +26.1 \\
Compound  & AICc SpIC & 28.3 25.0 & 21.8 19.4 & 26.9 24.0 & 0.767 0.796 &  9.8  2.5 & 212.2 170.7 & +27.5 +22.9 \\
\hline \hline
\multicolumn{9}{| c |} {\bf{Input $\daa=+10 \times 10^{-6}$}} \\
\hline
Turbulent & AICc SpIC & 31.0 27.9 & 23.6 21.8 & 27.9 24.8 & 0.763 0.791 &  9.5  2.7 & 197.6 160.8 & +9.62 +9.11 \\
Compound  & AICc SpIC & 23.5 21.8 & 18.5 17.5 & 22.3 20.8 & 0.762 0.785 &  9.5  1.6 & 182.4 148.9 & +9.66 +9.45 \\
\hline \hline
\multicolumn{9}{| c |} {\bf{Input $\daa=0$}} \\
\hline
Turbulent & AICc SpIC & {\color{blue} 31.6 28.8} & {\color{blue} 23.7 22.6} & {\color{blue} 27.7 25.5} & {\color{blue} 0.759 0.784} & {\color{blue} 9.4 2.7} & {\color{blue} 199.5 166.4} & {\color{blue} +0.22 +0.70} \\
Compound  & AICc SpIC & 25.6 23.1 & 20.2 18.8 & 24.5 22.3 & 0.762 0.783 &  8.7  2.0 & 191.9 158.2 & +2.33 +4.99 \\
\hline \hline
\multicolumn{9}{| c |} {\bf{Input $\daa=-10 \times 10^{-6}$}} \\
\hline
Turbulent & AICc SpIC & 32.1 29.0 & 24.6 22.8 & 28.0 25.6 & 0.760 0.785 &  9.7  2.5 & 204.5 167.6 & $-9.22$ $-8.10$ \\
Compound  & AICc SpIC & 27.7 24.7 & 22.3 20.4 & 26.6 24.0 & 0.765 0.789 &  8.1  1.5 & 203.2 166.4 & $-7.56$ $-4.67$ \\
\hline \hline
\multicolumn{9}{| c |} {\bf{M22 model $\daa=0$}} \\
\hline
\multicolumn{2}{| l |} {Turbulent}  & 41 & 27 & 26 & 0.795 & 0 & 177 & +2.2 \\
\hline
\end{tabular}
\caption{Mean number of quantities for each set of 25 {\sc ai-vpfit} models. The first three columns give mean numbers of absorption components in the models, for each atomic species. Where a column density iterated down to our minimum value of $\log N = 7.99$, the line is taken as a non-detection. $\langle$Int$\rangle$ is the mean number of interlopers i.e. unidentified absorption features that could not confidently be ascribed to one of the metals. $N_p$ is the total number of free parameters in a model (Equation \ref{eq:Np}). The two lines in blue font (AICc and SpIC, turbulent, input $\daa=0$) most directly compare with the M22 model.
\label{tab:n_components}
}
\end{table*}

The results of the 400 {\sc ai-vpfit} models are illustrated in Figures \ref{fig:turbulent_ICs} and \ref{fig:compound_ICs}, with numerical details in Tables \ref{tab:aicctb_results}, \ref{tab:spictb_results}, \ref{tab:aicccp_results}, and \ref{tab:spiccp_results}. Table \ref{tab:n_components} summarises the more detailed tables in the Appendix, and illustrates how easy it is to strongly bias results (rightmost column, $\daa$). 

The most important outcome of these {\sc ai-vpfit} calculations is that fixing $\daa$ during model construction creates severe bias. Every set of 25 calculations yields 25 final $\daa$ measurements that are all consistent with their input values. Even for a fairly extreme input fixed $\daa=+3 \times 10^{-5}$, the final measurements move only slightly away from their input value. Note that excellent fits are obtained for all input fixed $\daa$ settings (the mean normalised $\chi^2$ values are given in Table \ref{tab:n_components} and are all less than 0.8). This is not particularly surprising, given the large number of model parameters and their inter-dependency caused by line blending. 

Figure \ref{fig:all} illustrates one example turbulent model, with the spectral data. Figure \ref{fig:residuals} shows one turbulent model and one compound model, for two atomic transitions. Normalised residuals between the {\sc ai-vpfit} and M22 models are also illustrated, where the normalisation is done using the real spectral error array, thereby indicating model differences approximately in units of $\sigma$. Since each {\sc ai-vpfit} model is constructed differently, the number of model parameters varies from one model to the next.

\subsection{Number of parameters in the AI-VPFIT models}

The number of parameters in each {\sc ai-vpfit} model is given by
\be
\left.\begin{aligned}
N_p & = 5n_3 + 4n_2 + 3n_1 + 3n_{int} + n_T + 2n_{cont} + n_{zero} + n_{\alpha} \\
    & = 5n_3 + 4n_2 + 3n_1 + 3n_{int} + n_T + 18.
\end{aligned} \right.
\label{eq:Np}
\ee
$n_3$ is the number of metal velocity components comprising all three species (MgII, MgI, FeII), such that there are 5 parameters per velocity component, excluding temperature: $N_{MgII}, N_{FeII}, N_{MgI}, z, b_{turb}$. $n_2$ is the number of velocity components comprising only two species and $n_1$ is the number of components comprising only one species\footnote{Technical point: whilst our analysis requires all metal species to be present for each velocity component, sometimes this cannot be implemented because the spectral fitting ranges (which are those used in M22) do not cover exactly the same ranges in velocity space for all species.}. $n_{int}$ is the number of interlopers, each of which has 3 parameters, $N, z, b$. $n_T$ is the number of temperature parameters $n_T=n_3$ for compound broadening and $n_T=0$ for turbulent broadening. Each of the 8 spectral segments fitted has 2 continuum parameters, normalisation and slope\footnote{See {\sc vpfit} documentation \cite{web:VPFIT}} i.e. $2n_{cont} = 16$. One spectral segment in our models has a free zero level parameter (see {\sc vpfit} documentation), so $n_{zero}=1$. Each model has 1 $\daa$ parameter, so $n_{\alpha}=1$. The values of $N_p$ for each {\sc ai-vpfit} model are given in Tables \ref{tab:aicctb_results}, \ref{tab:spictb_results}, \ref{tab:aicccp_results}, and \ref{tab:spiccp_results}. Equation \ref{eq:Np} is specific to this particular modelling case and is not general.

{\sc ai-vpfit} {\it requires} components to be present in all 3 species. As discussed briefly in Section \ref{sec:Comparison}, this may seem like a fundamentally different approach between the two methods, but in fact it is not, because some of our components in FeII and MgI are very weak such that their column density parameters iterate down to our assigned lower bound of $\log N=7.99$. The apparently striking difference between {\sc ai-vpfit} models and the M22 model is thus more semantic than real, in this regard only. Table \ref{tab:n_components} gives the mean numbers of each species (MgII, MgI, and FeII) in each model. The mean value of the normalised $\chi^2$ for each type of fit is also shown. \\

\subsection{AI-VPFIT models -- turbulent broadening} \label{sec:turbulent}

The AICc and SpIC turbulent results show similarities although two differences are seen; the final $\daa$ SpIC measurements exhibit slightly more scatter than do the AICc results. This can be seen in Figure \ref{fig:turbulent_ICs} and numerically in the figure insets. Another notable difference is that the {\sc vpfit} error bars are smaller in most cases for SpIC than AICc. The explanation for both things is that SpIC requires fewer absorption components to achieve a similar goodness of fit (see the $\langle \chi^2_{\nu} \rangle$ values given in Table \ref{tab:n_components} and the numbers of components shown in Tables \ref{tab:aicctb_results} to \ref{tab:spiccp_results}). Fewer components translate (generally) to smaller parameter uncertainties (because there is less line blending) and fewer components create shallower false local minima. The two lines in blue font in Table \ref{tab:n_components} (AICc and SpIC, turbulent, input $\daa=0$) most directly compare with the M22 model. As noted earlier, and as is well-known, AICc has a tendency to over-fit. SpIC strikes a compromise between AICc and the Bayesian IC \cite{Webb2021}. The 25 AICc and 25 SpIC models have 199.5 and 166.4 free parameters respectively, compared with 177 for the M22 model.

\subsection{AI-VPFIT models -- compound broadening} \label{sec:compound}

The {\sc ai-vpfit} compound broadening models were computed primarily to explore whether the bias introduced by fixed $\daa$ is reduced or eliminated if a more physically appropriate line broadening mechanism is used. Figure \ref{fig:compound_ICs} shows that compound broadening models suffer just as badly from the bias caused by fixed $\daa$. 

The compound models illustrate smaller {\sc vpfit} error bars than the turbulent models. This is again to be expected because compound broadening generally required fewer absorption components to achieve a similar goodness of fit. The same argument then applies as given in the Subsection \ref{sec:turbulent}.

\section{Discussion} \label{sec:Discussion}

As discussed previously, the ``distortion blinding'' procedure of M22 (and other previous analyses in the literature) comprises two stages: (i) distort the wavelength scale such that an artificial $\daa \ne 0$ is added to the spectrum, and (ii) fix $\daa=0$ throughout the model construction process, releasing $\daa$ as a free parameter only {\it after} the final model has been obtained (allowing {\it all} parameters to vary in this last step). The calculations in the present paper have studied only the second effect.

This naturally raises the question: which of the procedures (i) and (ii) impose the strongest bias on the final measurement of $\daa$? Whilst we cannot answer this question quantitatively, we can do so qualitatively. The 400 {\sc ai-vpfit} models presented in the present paper strongly indicate that the fixed $\daa$ procedure (ii) imposes a catastrophic bias. Put simply, the presence of multiple minima in $\chi^2$-parameter space means that what goes in, comes out, producing a meaningless final measurement. By implication, process (i) will also bias the final measurement. However, because the added distortion is claimed to inflict an additional $\daa \lesssim 10^{-6}$, and because of the way in which process (ii) is seen to strongly bias results over far wider range in fixed $\daa$, it is reasonable to infer that process (ii) is probably the most damaging, at least in the case studied here.

The main conclusion of the calculations described in this paper is straightforward: ``distortion blinding'' is a mumpsimus and should now be abandoned; {\it neither of the processes (i) or (ii) should be employed in future varying alpha measurements}. Whilst we have not explicitly investigated process (i), we can infer from our results that this too has the capacity to bias the final result because it will create an incorrect initial model and a false local minimum. When trying to measure the fine structure constant in quasar absorption systems, the parameter $\daa$ should never be fixed at zero, or at any other value, during the model building process, other than at the very start of the model building process at which point only a single ``primary transition'' \citep{Lee2020AI-VPFIT} is involved, when redshift and $\daa$ are degenerate. As soon as additional transitions or species are included into the fit, $\daa$ must become a free parameter. If this is not done, $\daa$ is easily pushed into a false local minimum, from which no escape is likely. This implies that all previously published measurements of $\daa$ that have employed fixed $\daa=0$ should be repeated.

The conclusion above is fundamental to the way in which future measurements in this field of research should be made. Our results are derived from the analysis of only one quasar absorption system (although there is no reason to think that the $z_{abs}=1.15$ absorption system towards HE 0515$-$4414 is unique). A caveat is nevertheless that the conclusion expressed here relates to high signal to noise and high spectral resolution data (corresponding to ESPRESSO observations), both higher than the majority of previously published measurements. Further studies are required to expose similar effects for lower spectral resolution and lower signal to noise.

Since $\daa$ was fixed during the model building process for all {\sc ai-vpfit} calculations presented in this paper, {\it all} $\daa$ ``measurements'' reported here are biased and none should be considered as representative of the ``true'' value of $\daa$ in the $z_{abs}=1.15$ system towards HE 0515$-$4414. An unbiased analysis of this system will be presented in a separate paper.

\section*{Acknowledgements}
This research is based on observations collected at the European Southern Observatory under ESO programme 1102.A-0852. We are particularly grateful to the M22 team for making their reduced and co-added spectrum publicly available. We are also grateful for supercomputer time on OzSTAR at the Centre for Astrophysics and Supercomputing at Swinburne University of Technology and to the John Templeton Foundation for support in the early stages of this work. DM is supported by the INFN PD51 INDARK grant.

\section*{Data Availability}
The ESPRESSO spectra and associated files used for this analysis are available at \url{https://doi.org/10.5281/zenodo.5512490}. The 400 {\sc ai-vpfit} models will be available as online supplementary material on the MNRAS website, once accepted.

\bibliographystyle{mnras}
\bibliography{0515bias}

\appendix

\clearpage

\section{Tabulated AI-VPFIT results}

\begin{table*}
\centering
\begin{tabular}{llllr | lllr | lllr | lllr}
\hline
\multicolumn{5}{c |} {\bf{Input $\daa=+30 \times 10^{-6}$}} & \multicolumn{4}{c |} {\bf{Input $\daa=+10 \times 10^{-6}$}} & \multicolumn{4}{c |} {\bf{Input $\daa=0$}} & \multicolumn{4}{c} {\bf{Input $\daa=-10\times 10^{-6}$}} \\
\hline
 & Metals & Int & $N_p$ & $\daa$ & Metals & Int & $N_p$ & $\daa$ & Metals & Int & $N_p$ & $\daa$ & Metals & Int & $N_p$ & $\daa$  \\
\hline
 1  & 32 &  8 & 198 & 27.9 & 30 & 13 & 201 & 9.00 & 32 & 14 & 214 &  0.09 & 31 & 12 & 202 & -8.80 \\
 2  & 35 & 10 & 217 & 29.1 & 32 &  6 & 191 & 9.50 & 30 &  9 & 191 &  0.02 & 33 & 10 & 204 & -9.67 \\
 3  & 31 & 15 & 212 & 28.5 & 31 & 13 & 207 & 10.0 & 32 &  9 & 199 &  0.09 & 32 & 10 & 201 & -8.66 \\
 4  & 33 & 11 & 211 & 27.5 & 32 &  7 & 193 & 9.83 & 33 &  5 & 193 &  0.15 & 32 & 12 & 209 & -9.12 \\
 5  & 35 &  9 & 216 & 29.0 & 30 & 10 & 192 & 10.2 & 32 &  8 & 196 &  0.16 & 33 &  8 & 199 & -7.40 \\
 6  & 34 & 11 & 217 & 28.7 & 31 &  3 & 178 & 9.64 & 33 & 11 & 208 &  0.25 & 35 & 10 & 218 & -10.0 \\
 7  & 30 & 11 & 194 & 29.3 & 33 &  8 & 201 & 9.96 & 33 &  8 & 201 &  0.03 & 33 & 10 & 208 & -7.97 \\
 8  & 32 & 11 & 207 & 29.4 & 33 &  9 & 202 & 7.54 & 32 & 12 & 209 &  0.02 & 33 &  8 & 202 & -9.79 \\
 9  & 34 & 11 & 217 & 27.4 & 30 & 10 & 192 & 9.99 & 31 &  9 & 193 &  0.06 & 31 &  6 & 186 & -10.0 \\
 10 & 31 & 12 & 203 & 29.5 & 32 &  5 & 189 & 9.03 & 33 &  9 & 205 & -0.08 & 32 &  6 & 190 & -9.76 \\
 11 & 33 & 12 & 213 & 28.6 & 32 & 10 & 203 & 9.81 & 33 &  9 & 204 &  0.03 & 34 &  8 & 207 & -9.88 \\
 12 & 32 & 11 & 206 & 28.3 & 30 & 12 & 198 & 10.2 & 33 &  7 & 197 &  0.41 & 34 & 10 & 213 & -9.65 \\
 13 & 33 & 12 & 214 & 29.2 & 31 & 13 & 207 & 9.68 & 33 &  9 & 205 & -0.23 & 30 &  6 & 181 & -6.18 \\
 14 & 34 &  7 & 204 & 29.7 & 30 & 13 & 201 & 9.26 & 33 &  8 & 198 &  0.34 & 31 &  8 & 192 & -9.80 \\
 15 & 34 & 11 & 217 & 29.0 & 36 &  9 & 220 & 9.45 & 31 & 13 & 208 &  0.02 & 34 & 12 & 218 & -9.51 \\
 16 & 35 &  7 & 209 & 29.5 & 31 & 10 & 197 & 10.0 & 34 &  8 & 203 &  0.08 & 32 & 16 & 220 & -9.18 \\
 17 & 31 & 11 & 202 & 28.0 & 31 &  8 & 193 & 9.85 & 31 &  7 & 187 & -0.03 & 30 &  8 & 187 & -8.46 \\
 18 & 35 &  9 & 216 & 27.7 & 32 &  9 & 200 & 9.79 & 31 & 10 & 198 &  0.48 & 34 &  7 & 203 & -9.94 \\
 19 & 32 & 12 & 209 & 29.6 & 31 & 12 & 204 & 9.78 & 30 &  9 & 189 &  1.62 & 33 & 11 & 209 & -9.30 \\
 20 & 35 & 13 & 228 & 29.8 & 32 & 11 & 207 & 9.79 & 30 & 14 & 203 &  0.58 & 32 & 10 & 203 & -10.0 \\
 21 & 32 & 15 & 217 & 28.1 & 30 &  8 & 188 & 9.96 & 32 & 11 & 205 &  0.30 & 31 &  6 & 185 & -9.12 \\
 22 & 32 & 10 & 203 & 27.6 & 30 & 12 & 199 & 9.99 & 33 &  6 & 194 & -0.15 & 36 & 13 & 229 & -9.93 \\
 23 & 32 & 13 & 212 & 28.2 & 31 &  7 & 190 & 8.27 & 29 & 13 & 196 &  1.10 & 34 & 14 & 224 & -9.14 \\
 24 & 33 &  7 & 199 & 28.7 & 30 & 11 & 196 & 10.3 & 32 & 10 & 203 & -0.03 & 34 & 10 & 210 & -9.95 \\
 25 & 34 & 19 & 241 & 28.6 & 30 &  9 & 190 & 9.66 & 31 &  8 & 189 &  0.28 & 33 & 12 & 212 & -9.19 \\
 Means: & 33.0 & 11.1 & 211.3 & 28.7 & 31.2 & 9.5 & 197.6 & 9.62 & 31.9 & 9.4 & 199.5 & 0.22 & 32.7 & 9.7 & 204.5 & -9.22 \\
 \hline
\end{tabular}
\caption{Results from the AI Monte Carlo calculations for the 25 AICc turbulent models. ``Input $\daa=...$'' indicates the fixed value used to build the absorption system model. ``Int'' is the number of interlopers i.e. unidentified absorption features that could not confidently be ascribed to one of the metals. $N_p$ is the total number of free parameters in the model.
\label{tab:aicctb_results}
}
\end{table*}

\begin{table*}
\centering
\begin{tabular}{llllr | lllr | lllr | lllr}
\hline
\multicolumn{5}{c |} {\bf{Input $\daa=+30 \times 10^{-6}$}} & \multicolumn{4}{c |} {\bf{Input $\daa=+10 \times 10^{-6}$}} & \multicolumn{4}{c |} {\bf{Input $\daa=0$}} & \multicolumn{4}{c} {\bf{Input $\daa=-10\times 10^{-6}$}} \\
\hline
 & Metals & Int & $N_p$ & $\daa$ & Metals & Int & $N_p$ & $\daa$ & Metals & Int & $N_p$ & $\daa$ & Metals & Int & $N_p$ & $\daa$  \\
\hline
 1  & 28 &  4 & 166 & 23.5 & 27 & 1 & 151 & 10.2 & 30 & 3 & 170 &  0.34 & 28 & 3 & 162 & -7.87 \\
 2  & 26 &  4 & 155 & 26.3 & 28 & 2 & 158 & 7.70 & 28 & 1 & 155 &  1.75 & 28 & 3 & 162 & -7.55 \\
 3  & 31 &  1 & 171 & 28.1 & 32 & 9 & 201 & 10.0 & 28 & 1 & 154 &  1.00 & 28 & 6 & 170 & -8.29 \\
 4  & 30 &  4 & 176 & 28.5 & 26 & 2 & 148 & 10.2 & 26 & 1 & 145 &  2.53 & 29 & 3 & 165 & -8.42 \\
 5  & 28 &  1 & 157 & 21.2 & 25 & 2 & 143 & 9.98 & 29 & 1 & 160 &  0.25 & 29 & 1 & 161 & -8.22 \\
 6  & 31 &  3 & 179 & 29.6 & 28 & 1 & 156 & 8.11 & 28 & 3 & 161 &  0.30 & 32 & 6 & 189 & -9.96 \\
 7  & 31 &  6 & 187 & 27.5 & 30 & 1 & 167 & 9.08 & 28 & 6 & 171 &  0.94 & 31 & 2 & 173 & -7.06 \\
 8  & 28 &  2 & 159 & 23.8 & 27 & 3 & 157 & 9.18 & 30 & 5 & 178 &  0.02 & 28 & 1 & 155 & -8.50 \\
 9  & 30 &  6 & 181 & 28.3 & 26 & 2 & 149 & 8.36 & 29 & 2 & 163 &  0.61 & 30 & 2 & 167 & -6.64 \\
 10 & 30 &  2 & 170 & 28.4 & 27 & 3 & 157 & 10.1 & 27 & 3 & 156 &  1.33 & 32 & 2 & 178 & -7.81 \\
 11 & 29 &  1 & 162 & 25.3 & 30 & 4 & 174 & 9.97 & 31 & 1 & 170 &  0.90 & 29 & 1 & 160 & -6.68 \\
 12 & 32 &  1 & 176 & 23.7 & 28 & 5 & 167 & 8.46 & 29 & 2 & 162 &  0.54 & 29 & 2 & 164 & -4.71 \\
 13 & 33 &  5 & 194 & 27.4 & 28 & 2 & 159 & 8.79 & 32 & 4 & 185 & -0.22 & 30 & 7 & 183 & -9.33 \\
 14 & 26 &  3 & 152 & 23.9 & 27 & 4 & 160 & 8.67 & 31 & 1 & 169 & -0.03 & 30 & 4 & 174 & -8.60 \\
 15 & 28 &  7 & 173 & 26.5 & 31 & 2 & 174 & 9.51 & 27 & 8 & 173 &  0.12 & 29 & 3 & 167 & -7.68 \\
 16 & 28 &  1 & 158 & 25.0 & 28 & 2 & 159 & 10.6 & 30 & 2 & 167 & -0.02 & 31 & 2 & 173 & -9.45 \\
 17 & 28 &  2 & 160 & 24.0 & 26 & 1 & 145 & 8.77 & 31 & 2 & 174 &  0.33 & 32 & 2 & 178 & -8.47 \\
 18 & 31 &  4 & 180 & 25.7 & 30 & 3 & 171 & 8.93 & 31 & 3 & 176 & -0.00 & 29 & 2 & 163 & -8.35 \\
 19 & 30 &  0 & 165 & 28.1 & 31 & 1 & 171 & 10.1 & 29 & 1 & 162 &  0.29 & 30 & 1 & 165 & -6.31 \\
 20 & 28 &  0 & 153 & 26.3 & 28 & 2 & 158 & 9.37 & 29 & 1 & 160 &  2.17 & 31 & 2 & 173 & -9.23 \\
 21 & 29 &  2 & 164 & 27.5 & 27 & 6 & 166 & 7.99 & 27 & 4 & 160 &  0.99 & 30 & 1 & 164 & -10.0 \\
 22 & 27 &  2 & 154 & 23.6 & 28 & 1 & 155 & 9.24 & 29 & 1 & 162 &  2.14 & 28 & 1 & 157 & -9.68 \\
 23 & 27 & 12 & 186 & 27.2 & 27 & 1 & 151 & 9.55 & 31 & 4 & 180 & -0.04 & 30 & 2 & 168 & -6.73 \\
 24 & 30 &  0 & 163 & 26.9 & 27 & 6 & 167 & 8.11 & 31 & 4 & 180 &  0.03 & 29 & 2 & 164 & -10.0 \\
 25 & 29 &  2 & 166 & 25.4 & 28 & 1 & 156 & 6.67 & 29 & 3 & 166 &  1.31 & 29 & 1 & 161 & -6.92 \\
 Means: & 29.1 & 3.0 & 168.3 & 26.1 & 28.0 & 2.7 & 160.8 & 9.11 & 29.2 & 2.7 & 166.4 & 0.70 & 29.6 & 2.5 & 167.6 & -8.10 \\
 \hline
\end{tabular}
\caption{As Table \ref{tab:aicctb_results} but for SpIC turbulent results.
\label{tab:spictb_results}
}
\end{table*}

\begin{table*}
\centering
\begin{tabular}{llllr | lllr | lllr | lllr}
\hline
\multicolumn{5}{c |} {\bf{Input $\daa=+30 \times 10^{-6}$}} & \multicolumn{4}{c |} {\bf{Input $\daa=+10 \times 10^{-6}$}} & \multicolumn{4}{c |} {\bf{Input $\daa=0$}} & \multicolumn{4}{c} {\bf{Input $\daa=-10\times 10^{-6}$}} \\
\hline
 & Metals & Int & $N_p$ & $\daa$ & Metals & Int & $N_p$ & $\daa$ & Metals & Int & $N_p$ & $\daa$ & Metals & Int & $N_p$ & $\daa$  \\
\hline
 1  & 28 &  8 & 204 & 28.3 & 23 & 10 & 182 & 9.17 & 28 & 11 & 212 & 0.01 & 28 &13 & 219 & -8.00 \\
 2  & 30 & 12 & 228 & 28.9 & 24 & 13 & 198 & 9.47 & 25 &  7 & 185 & 3.88 & 26 & 6 & 189 & -3.85 \\
 3  & 26 &  8 & 195 & 26.1 & 23 & 11 & 184 & 9.83 & 25 &  7 & 185 & 3.29 & 27 & 9 & 203 & -8.13 \\
 4  & 29 & 14 & 226 & 27.7 & 24 & 12 & 195 & 9.25 & 25 &  6 & 181 & 1.77 & 27 &10 & 206 & -8.39 \\
 5  & 30 &  9 & 218 & 27.2 & 25 & 10 & 192 & 9.45 & 24 & 10 & 189 & 5.80 & 29 & 6 & 200 & -8.73 \\
 6  & 27 &  9 & 204 & 24.5 & 22 & 10 & 177 & 9.73 & 26 &  6 & 186 & 0.11 & 31 &10 & 227 & -9.20 \\
 7  & 29 &  7 & 210 & 24.5 & 23 & 11 & 183 & 9.67 & 27 &  6 & 189 & 0.96 & 27 & 9 & 204 & -8.06 \\
 8  & 29 &  8 & 209 & 28.6 & 23 & 10 & 180 & 9.81 & 25 & 10 & 190 & 3.85 & 26 & 9 & 195 & -5.06 \\
 9  & 26 & 12 & 204 & 27.7 & 24 &  6 & 171 & 9.56 & 25 & 11 & 198 & 3.17 & 27 & 7 & 195 & -7.82 \\
 10 & 26 & 10 & 201 & 27.4 & 25 &  8 & 188 & 9.56 & 27 & 10 & 206 & 1.24 & 27 & 9 & 204 & -8.45 \\
 11 & 30 &  9 & 220 & 28.2 & 23 & 10 & 180 & 9.86 & 25 & 10 & 191 & 4.14 & 27 & 8 & 201 & -8.36 \\
 12 & 29 &  6 & 207 & 28.8 & 25 & 10 & 193 & 9.99 & 25 &  8 & 187 & 1.86 & 26 &11 & 201 & -7.37 \\
 13 & 28 & 12 & 216 & 25.1 & 25 & 10 & 189 & 9.85 & 25 & 12 & 200 & 3.38 & 26 & 8 & 192 & -6.56 \\
 14 & 26 &  9 & 197 & 26.5 & 25 &  9 & 192 & 9.70 & 26 &  8 & 189 & 1.86 & 28 & 5 & 194 & -7.77 \\
 15 & 31 &  6 & 216 & 29.4 & 23 & 10 & 182 & 9.59 & 26 &  7 & 188 & 1.50 & 28 & 9 & 207 & -4.33 \\
 16 & 27 & 11 & 207 & 28.6 & 23 & 10 & 180 & 9.88 & 27 &  9 & 202 & 2.04 & 29 & 5 & 204 & -7.78 \\
 17 & 31 & 10 & 229 & 28.5 & 24 &  9 & 180 & 9.86 & 27 &  6 & 190 & 1.92 & 29 & 7 & 208 & -8.86 \\
 18 & 29 & 12 & 222 & 28.8 & 23 &  8 & 176 & 9.75 & 25 & 10 & 192 & 3.29 & 31 & 7 & 220 & -8.47 \\
 19 & 27 & 11 & 210 & 28.8 & 23 &  9 & 177 & 9.66 & 25 &  7 & 185 & 1.90 & 27 & 8 & 200 & -6.18 \\
 20 & 31 &  6 & 218 & 28.7 & 23 & 11 & 183 & 10.0 & 26 &  8 & 192 & 1.64 & 28 & 8 & 204 & -6.78 \\
 21 & 28 & 15 & 225 & 28.3 & 22 &  9 & 172 & 9.44 & 25 & 11 & 195 & 3.45 & 27 & 8 & 200 & -5.53 \\
 22 & 28 &  8 & 204 & 26.5 & 24 &  8 & 180 & 10.0 & 25 & 11 & 195 & 2.58 & 26 & 9 & 198 & -7.97 \\
 23 & 29 & 13 & 225 & 28.7 & 23 &  3 & 159 & 9.99 & 25 &  9 & 189 & 1.88 & 27 & 6 & 191 & -9.02 \\
 24 & 27 & 13 & 214 & 26.4 & 23 &  9 & 177 & 8.75 & 26 &  9 & 192 & 1.05 & 29 & 6 & 205 & -8.62 \\
 25 & 27 &  6 & 195 & 24.9 & 23 & 12 & 189 & 9.75 & 26 &  8 & 190 & 1.71 & 29 & 9 & 213 & -9.60 \\
 Means: & 28.3 & 9.8 & 212.2 & 27.5 & 23.5 & 9.5 & 182.4 & 9.66 & 25.6 & 8.7 & 191.9 & 2.33 & 27.7 & 8.1 & 203.2 & -7.56 \\
 \hline
\end{tabular}
\caption{As Table \ref{tab:aicctb_results} but for the 25 AICc compound models.
\label{tab:aicccp_results}
}
\end{table*}

\begin{table*}
\centering
\begin{tabular}{llllr | lllr | lllr | lllr}
\hline
\multicolumn{5}{c |} {\bf{Input $\daa=+30 \times 10^{-6}$}} & \multicolumn{4}{c |} {\bf{Input $\daa=+10 \times 10^{-6}$}} & \multicolumn{4}{c |} {\bf{Input $\daa=0$}} & \multicolumn{4}{c} {\bf{Input $\daa=-10\times 10^{-6}$}} \\
\hline
 & Metals & Int & $N_p$ & $\daa$ & Metals & Int & $N_p$ & $\daa$ & Metals & Int & $N_p$ & $\daa$ & Metals & Int & $N_p$ & $\daa$  \\
\hline
 1  & 24 & 2 & 159 & 19.6 & 21 & 1 & 143 & 9.64 & 22 & 2 & 150 & 7.10 & 24 & 1 & 161 & -3.71 \\
 2  & 25 & 2 & 168 & 23.2 & 22 & 1 & 150 & 9.58 & 23 & 3 & 161 & 2.97 & 27 & 2 & 182 & -7.15 \\
 3  & 26 & 4 & 182 & 23.2 & 22 & 3 & 156 & 9.66 & 21 & 3 & 148 & 6.49 & 26 & 1 & 171 & -1.29 \\
 4  & 26 & 2 & 177 & 22.5 & 23 & 2 & 159 & 9.43 & 22 & 1 & 148 & 7.83 & 25 & 2 & 169 & -5.20 \\
 5  & 26 & 1 & 169 & 22.8 & 22 & 1 & 148 & 9.42 & 22 & 2 & 152 & 4.92 & 24 & 0 & 156 & -6.25 \\
 6  & 24 & 3 & 169 & 26.2 & 23 & 1 & 151 & 9.45 & 24 & 4 & 168 & 3.42 & 24 & 1 & 159 & -7.36 \\
 7  & 25 & 3 & 174 & 21.2 & 22 & 1 & 147 & 9.18 & 23 & 1 & 156 & 5.57 & 23 & 2 & 156 & -8.04 \\
 8  & 26 & 3 & 176 & 23.3 & 22 & 3 & 156 & 9.65 & 24 & 2 & 165 & 6.02 & 24 & 1 & 159 & -6.83 \\
 9  & 23 & 3 & 159 & 18.3 & 22 & 1 & 147 & 9.64 & 24 & 1 & 161 & 4.70 & 24 & 2 & 165 & -4.31 \\
 10 & 25 & 3 & 174 & 23.2 & 23 & 1 & 153 & 9.19 & 25 & 1 & 168 & 3.39 & 23 & 2 & 159 & -8.06 \\
 11 & 26 & 4 & 183 & 28.0 & 21 & 1 & 144 & 9.71 & 24 & 1 & 162 & 4.27 & 23 & 1 & 156 &  0.19 \\
 12 & 26 & 1 & 174 & 22.2 & 22 & 1 & 150 & 9.62 & 24 & 3 & 165 & 4.67 & 25 & 1 & 165 & -4.46 \\
 13 & 25 & 3 & 171 & 23.2 & 22 & 2 & 153 & 9.67 & 24 & 3 & 168 & 3.68 & 23 & 1 & 156 &  0.64 \\
 14 & 24 & 2 & 161 & 21.2 & 21 & 1 & 141 & 9.31 & 23 & 4 & 164 & 6.21 & 26 & 1 & 171 & -7.24 \\
 15 & 24 & 4 & 171 & 22.5 & 21 & 1 & 143 & 9.50 & 23 & 2 & 156 & 6.23 & 25 & 3 & 174 & -3.25 \\
 16 & 24 & 5 & 171 & 23.5 & 22 & 4 & 156 & 9.48 & 24 & 1 & 162 & 3.81 & 24 & 2 & 165 &  0.69 \\
 17 & 24 & 3 & 166 & 21.9 & 21 & 2 & 144 & 9.53 & 25 & 1 & 166 & 5.33 & 24 & 3 & 167 & -4.58 \\
 18 & 26 & 2 & 172 & 22.8 & 22 & 1 & 150 & 9.91 & 22 & 1 & 147 & 5.93 & 26 & 1 & 174 & -5.53 \\
 19 & 26 & 2 & 175 & 25.5 & 21 & 2 & 144 & 9.60 & 23 & 2 & 154 & 5.95 & 27 & 3 & 186 & -6.31 \\
 20 & 22 & 2 & 149 & 19.2 & 22 & 3 & 153 & 9.09 & 21 & 3 & 148 & 2.30 & 25 & 1 & 168 &  1.61 \\
 21 & 25 & 1 & 168 & 21.3 & 21 & 1 & 141 & 9.44 & 22 & 1 & 149 & 6.80 & 26 & 1 & 174 & -6.30 \\
 22 & 26 & 1 & 171 & 23.1 & 22 & 2 & 152 & 9.59 & 23 & 1 & 156 & 3.63 & 25 & 1 & 165 & -7.99 \\
 23 & 27 & 2 & 183 & 27.0 & 22 & 1 & 150 & 9.48 & 22 & 1 & 150 & 5.69 & 24 & 1 & 162 & -3.50 \\
 24 & 25 & 4 & 175 & 25.8 & 21 & 1 & 141 & 8.85 & 24 & 4 & 168 & 3.96 & 26 & 3 & 178 & -8.56 \\
 25 & 26 & 1 & 170 & 21.3 & 22 & 1 & 150 & 9.71 & 24 & 2 & 162 & 3.94 & 25 & 1 & 163 & -3.88 \\
 Means: & 25.0 & 2.5 & 170.7 & 22.9 & 21.8 & 1.6 & 148.9 & 9.49 & 23.1 & 2.0 & 158.2 & 4.99 & 24.7 & 1.5 & 166.4 & -4.67 \\
 \hline
\end{tabular}
\caption{As Table \ref{tab:aicctb_results} but for the 25 SpIC compound models.
\label{tab:spiccp_results}
}
\end{table*}

\clearpage

\section{Example model fits}

The following three figures are the same as Figure \ref{fig:all} but $\daa = -10, +10, +30 \times10^{-6}$. 

\begin{figure*}
\centering
\includegraphics[width=0.9\linewidth]{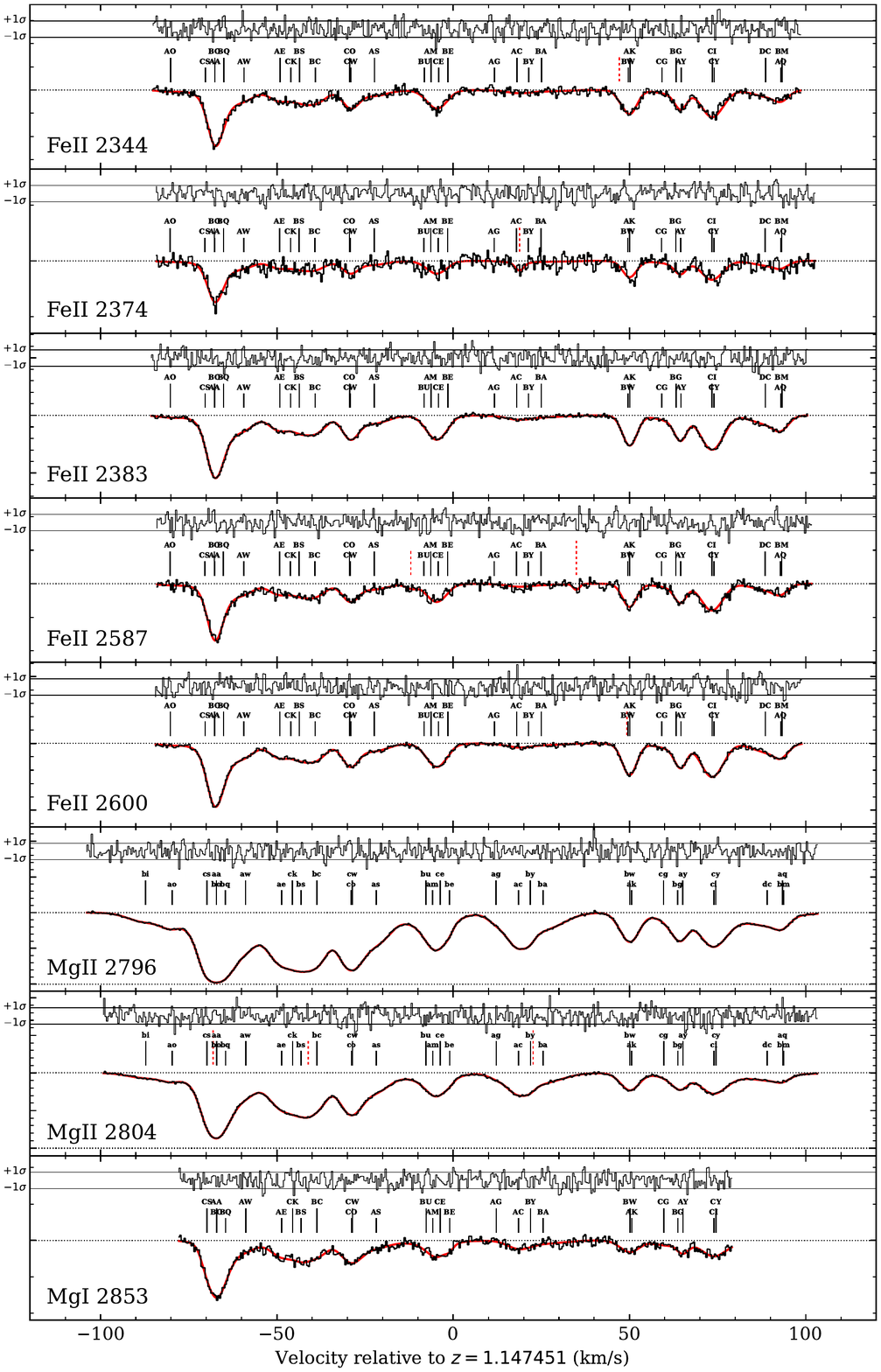}
\caption{The same as Figure \ref{fig:all} but for $\daa=+30\times10^{-6}$ and model 1 in the first column of Table \ref{tab:aicctb_results}. This model was chosen because its $N_{p}=198$ is the lowest in that column.
\label{fig:model_p30} 
}
\end{figure*}

\begin{figure*}
\centering
\includegraphics[width=0.9\linewidth]{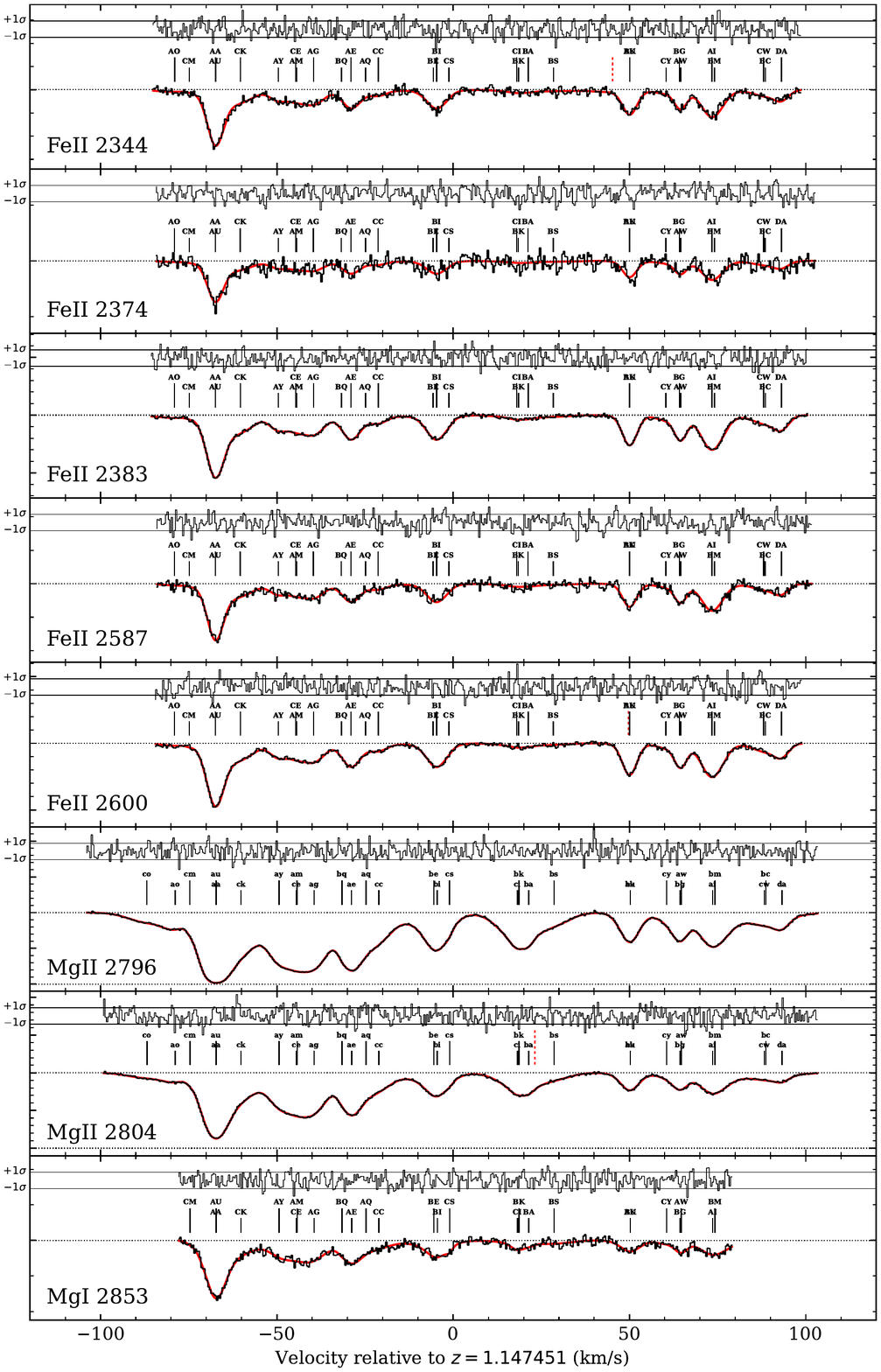}
\caption{The same as Figure \ref{fig:all} but for $\daa=+10\times10^{-6}$ and model 5 in the second column of Table \ref{tab:aicctb_results}. This model was chosen because its $N_{p}=178$ is the closest to the that of M22 model (177).
\label{fig:modelp10} 
}
\end{figure*}

\begin{figure*}
\centering
\includegraphics[width=0.9\linewidth]{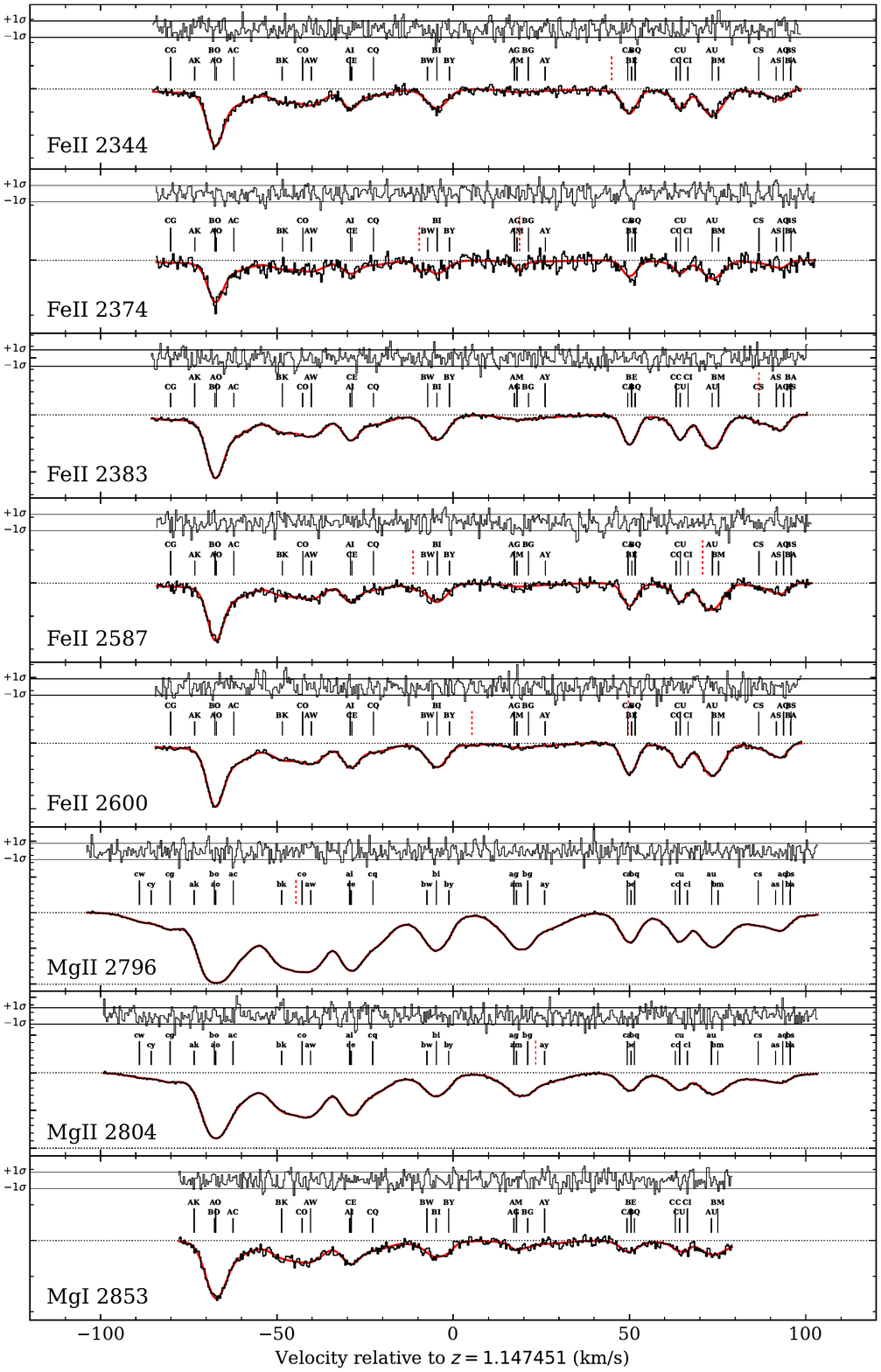}
\caption{The same as Figure \ref{fig:all} but for $\daa=-10\times10^{-6}$ and model 2 in the fourth column of Table \ref{tab:aicctb_results}. This model was chosen because its $N_{p}=204$ is the closest to the mean of that column (204.5).
\label{fig:modelm10} 
}
\end{figure*}

\bsp
\label{lastpage}

\end{document}